\documentclass[pra, reprint, twocolumn, superscriptaddress, amsfonts, amssymb, amsmath]{revtex4-2}
\usepackage[english]{babel}

\usepackage{graphicx}

\graphicspath{{./Images/},{./ImagesAppendix/},
{./images/},{./imagesAppendix/}}

\usepackage{physics}
\usepackage{bm}
\usepackage{braket}
\usepackage{pgfplots}
\usepackage{array}
\usepackage{makecell}
\usepackage{amsmath}
\usepackage{tikz}
\usetikzlibrary{positioning, arrows.meta}
\usepackage{subfigure}
\usepackage{commath}
\usepackage{graphicx,bm}
\usepackage{verbatim}
\usepackage{flafter}
\usepackage{float}
\usepackage{varioref}

\usepackage{mathtools}

\usepackage{color}
\usepackage[colorlinks,citecolor=darkBlue,linkcolor=darkBlue,
urlcolor=blue,hyperindex]{hyperref}
\definecolor{darkBlue}{rgb}{0.08, 0.13, 0.4}

\definecolor{THc}{rgb}{0.9,0.3,0.2}

\newcommand{\canc}[1]{}

\begin{document}

\title{Predicting fermionic densities using a Projected Quantum Kernel method}
\author{F. Perciavalle}
\affiliation{Dipartimento di Fisica, Universit\`a della Calabria, 87036 Arcavacata di Rende (CS), Italy}
\affiliation{INFN--Gruppo collegato di Cosenza}
\author{F.Plastina}
\affiliation{Dipartimento di Fisica, Universit\`a della Calabria, 87036 Arcavacata di Rende (CS), Italy}
\affiliation{INFN--Gruppo collegato di Cosenza}
\author{M.Pisarra}
\affiliation{Dipartimento di Fisica, Universit\`a della Calabria, 87036 Arcavacata di Rende (CS), Italy}
\affiliation{INFN--Gruppo collegato di Cosenza}\author{N. Lo Gullo}
\affiliation{Dipartimento di Fisica, Universit\`a della Calabria, 87036 Arcavacata di Rende (CS), Italy}
\affiliation{INFN--Gruppo collegato di Cosenza}

\date{\today}

\begin{abstract}
We use a support vector regressor based on a projected quantum kernel method to predict the density structure of 1D fermionic systems of interest in quantum chemistry and quantum matter. The kernel is built on
with the observables of a quantum reservoir implementable with interacting Rydberg atoms. Training and test data of the fermionic system are generated using a Density Functional Theory approach. We test the performance of the method for several Hamiltonian parameters, finding a general common behavior of the error as a function of measurement time. At sufficiently large measurement times, we find that the method outperforms the classical linear kernel method and can be competitive with the radial basis function method.
\end{abstract}

\maketitle

\section{Introduction}
Support vector machines (SVMs) are supervised learning models widely used for classification (SVC) and regression (SVR) of data; their operation is based on finding an optimal hyperplane that separates or interpolates the data~\cite{hearst1998support,kecman2005support,scikit2011pedregosa,pml1Book,muller2001introduction}. In many cases, the relationship between input and output data is nonlinear, so a hyperplane constructed in the input data space alone is insufficient to capture the underlying pattern. The problem is, then, addressed by the kernel trick, a method that involves mapping the input data into a higher-dimensional space, facilitating the search for the aforementioned hyperplane.

SVMs have proven effective in addressing complex problems in the physical sciences~\cite{carleo2019machine}, for instance by circumventing demanding numerical methods like Density Functional Theory (DFT)~\cite{brockherde2017bypassing, kohn1965self, kohn1999nobel, hohenberg1964inhomogeneous, giuliani2005quantum}. DFT is a theoretical approach to investigate the structure of materials and molecules whose aim is to express the problem in terms of the electronic density, without accessing the exact many-body wave-function of the system. The density distribution is obtained by numerically solving the self-consistent Kohn-Sham (KS) equations~\cite{kohn1965self}. Determining the density distribution by bypassing the KS equations can potentially speed up the calculation, and SVMs are a possible tool for doing so~\cite{brockherde2017bypassing}.

Machine Learning (ML) is a resource for quantum physics and vice versa. Indeed, with the emergence of Noisy Intermediate-Scale Quantum (NISQ) devices~\cite{bharti2022noisy}, there has been a growing interest in exploiting quantum resources to improve the performance of classical ML methods. This has led to the development of the field of Quantum Machine Learning (QML)~\cite{biamonte2017quantum}. For instance, the intricate dynamics of quantum systems can be leveraged to uncover the complex relationships between input and output data, as demonstrated by Quantum Extreme Learning and Quantum Reservoir Computing methods~\cite{fujii2016harnessing,fujii2019boosting,settino2024memory,mujal2021opportunities, martine2021dynamical, zia2025quantum, delorenzis2024harnessing, beaulieu2024robust, Lomonaco2024quantum}. Moreover, in the context of SVM, quantum physics underlies the concept of Quantum Kernels (QK) and Projected Quantum Kernels (PQK)~\cite{huang2021power,schnabel2024quantum, haug2023quantum, blank2020quantum, mengoni2019kernel,huang2021power,damore2024projected}. In the former, the Hilbert space of a quantum system is used to expand the dimension of the input data space; in the latter, after mapping to Hilbert space, a new mapping is made to a classical space by, for example, measuring physical observables. The mapping can be performed by encoding the input data in the Hamiltonian parameters of a controllable quantum system, identified as quantum reservoir, which is left to evolve in time before performing the measurement of a certain set of observables. The vector containing the observables defines the mapping of input data in a enlarged space and can then be used for building a new kernel, that can be identified as a PQK. In Ref.~\cite{kornjaca2024large}, this approach has been applied to many different tasks using a Rydberg atom system as a reservoir. 

\begin{figure*}[htbp]
\centering
\includegraphics[width=\linewidth]{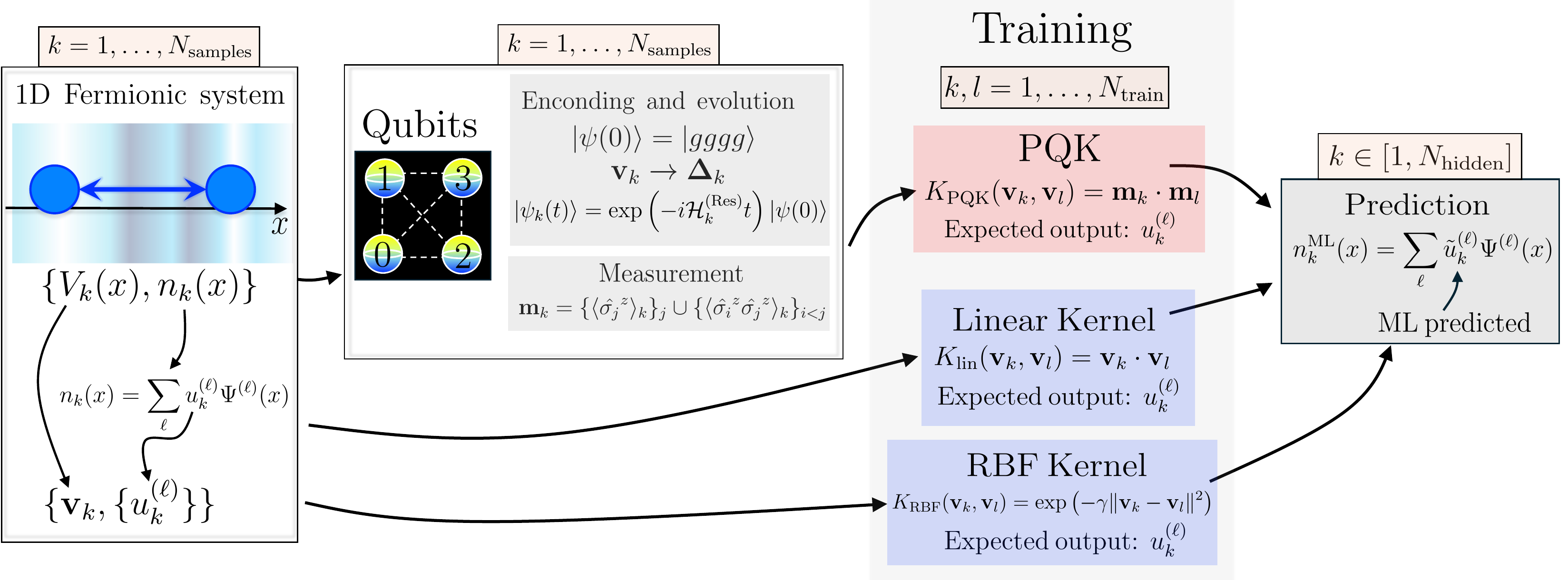}
\caption{Sketch of the method: the two-particle fermionic ground state density is found using the KS equations. For any configuration of the potential features $\bm{\mathrm{v}}_k$, a density profile codified by $\{u_k^{(\ell)}\}_{\ell}$ is saved and then used for training. Before training, the potential features are encoded in the detunings of the reservoir Hamiltonian, whose evolution is used to enlarge the feature space. At a specific measurement time a vector containing magnetizations and correlations is defined. The latter is used to build the kernel $K_{\rm PQK}$, while linear and RBF kernels are built directly by using the potential features; the coefficients $\{u_k^{(\ell)}\}_{\ell}$ are the expected outputs. Given a potential features configuration $\bm{\mathrm{v}}_k$, we obtain trained SVR machines that predict the coefficients $\tilde{u}_k^{(\ell)}$, separately for any $\ell$. The coefficients are then used to reconstruct the ML-predicted density profile $n_k^{\rm ML}(x)$.}
\label{fig:sketch}
\end{figure*}
In this paper, we use the aforementioned approach with a reservoir composed of few interacting qubits. The task is to predict the fermionic densities of specific 1D quantum systems using information about the external potential to which the fermions are subjected, with the aim of bypassing the KS equations~\cite{brockherde2017bypassing}. The systems analyzed are of interest in quantum chemistry and quantum matter. In fact, simple one-dimensional molecules have been shown to behave similarly to their real three-dimensional counterparts~\cite{wagner2012reference}, making them a theoretical laboratory for real systems~\cite{stoudenmire2012one,loos2015chemistry, ball2015chem1D, baker2015one, li2021exact, li2016pure,li2015understanding, snyder2013orbital, gedeon2021machine}. Other interesting models include 1D fermionic systems in multiple external well potentials, that have been recently studied using ML approaches~\cite{sarkar2025physics} and are basic models for quantum devices such as superlattices~\cite{esaki1970superlattice, harrison2005quantum, kasap2017handbook} and quantum cascade lasers~\cite{faist1994quantum, jirauschek2014modeling}.

In the context of 1D Chemistry, we consider two paradigmatic cases of study: two fermions in an external potential provided by two hydrogen nuclei, i.e. a one-dimensional $H_2$ molecule, and two fermions in a triple-well external potential. In both of the problems, data are generated by employing a DFT approach that makes use of the 1D Local Density Approximation (LDA), proposed in Ref.~\cite{Entwistle2016local}. Beside the form of the external potential, the two problems differ on the number of potential features that are used for training. 
The features are specifically encoded in the detunings of the reservoir Hamiltonian, following the so-called local encoding~\cite{kornjaca2024large}. Our aim is to analyze the performance of the method and compare it with those of classical methods. Its effectiveness can serve as a starting point for using it to explore the structure of different 1D, 2D and 3D quantum systems, including molecules.

The paper is organized as follows. In section~\ref{sec:methods} we describe the model whose density we predict and the method used to do so. In section~\ref{sec:results}, we report our results by analyzing the performance of the method in two different problems: the 1D $H_2$ molecule, in subsection~\ref{subsec:H2}, and the triple-well potential, in subsection~\ref{subsec:Triple}. In section~\ref{sec:conclusions} we recall the key points of the method, summarize our results, and provide possible perspectives. Finally, the appendices contain calculations and results that support the findings of the main text.

\section{Model and Methods}
\label{sec:methods}
We propose a Support Vector Regressor (SVR) to predict the spatial density 
of interacting quantum particles in an external potential.
As case of study, we consider a system composed of two interacting fermions subjected to an external potential. The Hamiltonian of the system is 
\begin{equation}
\mathcal{H}^{(2\rm F)} = \sum_{i=1,2}\left[T_i + V(x_i)\right] + U(|x_1 - x_2|),
\label{eq:H_ferm}
\end{equation}
where $T_i$ is the kinetic energy, $V(x_i)$ is the external potential to which the $i$-th fermion is subjected to and $U(|x_1-x_2|)=(|x_1-x_2|+1)^{-1}$ is the interaction modeled as a softened Coulomb repulsion~\cite{Entwistle2016local, Hodgson2013exact}. 
We follow the ML-HK (Machine Learning - Hohenberg Kohn) mapping approach proposed in Ref.~\cite{brockherde2017bypassing}, according to which, the external potential is mapped into the density distribution by using ML techniques, in this work we use a SVR. Since the problem is highly non-linear, we resort to the kernel trick, where the latter is implemented via the dynamics of a quantum system.

Before implementing the SVR, we prepare data for training and testing machine performance.
Specifically, we consider different configurations of the external potential $V_k(x)$ labeled with $k=1,\ldots,N_{\rm samples}$ which we thereafter call samples. The samples are divided into $N_{\rm train}$ training samples and $N_{\rm hidden}$ hidden samples such that $N_{\rm samples}=N_{\rm train}+N_{\rm hidden}$. The training samples are used to train the machine, whereas the hidden samples are used to test its performance. At any $k$, the features of the potential are encoded in a finite-dimensional input vector $\bm{\mathrm{v}}_k$. The density distribution is found using a DFT~\cite{hohenberg1964inhomogeneous,kohn1999nobel,burke2012perspective} approach, based on the solution of the KS equation~\cite{kohn1965self}. The latter is a one-particle self-consistent equation in which potential and interaction effects are embedded in the KS potential
\begin{equation}
V_{\rm KS,k}(x)=V_k(x)+V_{\rm H}(x) + V_{\rm xc}(x), 
\end{equation}
where $V_{\rm H}(x)=\int dx' \frac{n(x')}{|x-x'|+1}$ is the Hartree potential and $V_{\rm xc}(x)=\left( - 1.19 + 1.77 n(x) - 1.37 n(x)^2\right) n(x)^{0.604}$ is the exchange-correlation potential in LDA~\cite{Entwistle2016local}, see Appendix~\ref{app:KS} for further details. 

The target variables on which the SVR is trained are not the grid points of the density distribution, but the coefficients of the expansion of the latter in a chosen basis. Quantitatively, this means that we expand each sample $k$ density profile $n_k(x)$ in a given basis $\{\Psi^{(\ell)}(x)\}_{\ell}$:
\begin{equation}
n_k(x)=\sum_{\ell}u_k^{(\ell)}\Psi^{(\ell)}(x).
\label{eq:expansion}
\end{equation}
The basis is chosen in such a way that the density profile of the system can be reproduced with a finite number of elements within a chosen error. In this spirit, we truncate the expansion
\begin{equation}
n_k^{(\rm approx)}(x)=\sum_{\ell=1}^{N_{\rm trunc}}u_k^{(\ell)}\Psi^{(\ell)}(x)\approx n_k(x)   
\label{eq:approx}
\end{equation}
and we collect the coefficients $\{ u_k^{(\ell)} \}_{\ell=1}^{N_{\rm trunc}}$, which are the target variables of the ML. The ML is performed through SVRs on each $\ell$ separately, using a kernel function $K(\bm{\mathrm{v}}_{k},\bm{\mathrm{v}}_{s})$ that depends on the input data $\bm{\mathrm{v}}_k$.
In our work we considered and compared the performances of three different kernels: i) Linear (lin); ii) Radial Basis Function (RBF); iii) PQK built with physical observables. The linear and RBF kernels are standard ones, and correspond to the case in which one considers the same number of features as input and an infinite one respectively, see Appendix~\ref{app:SVR} for more details. 

We now describe our implementation of the PQK. 

\begin{figure*}[t!]
\centering
\includegraphics[width=\linewidth]{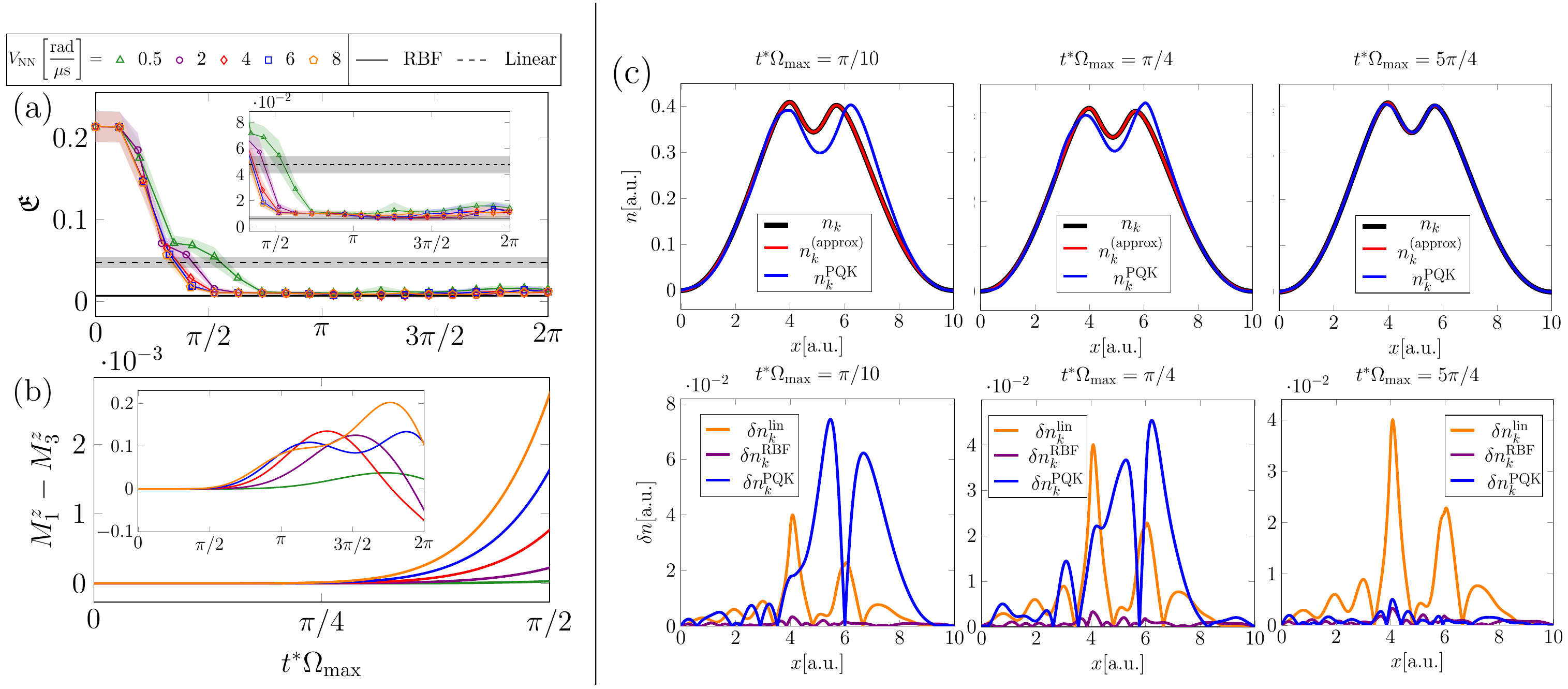}
\caption{Comparison between classical and PQK SVR methods for $H_2$. Panel (a): error of PQK method in function of the measurement time and compared with those of Linear and RBF methods. Different values of the nearest neighbor interaction $V_{\rm NN}$ are considered. The time is rescaled with the maximum frequency $\Omega_{\rm max}$. The parameters of the reservoir are set to $\Omega_{\rm glob}=5$ rad/$\mu$s, $\Delta_{\rm glob}=0$, $\Delta_{\rm loc}=-3.5$ rad/$\mu$s. The homogeneous detuning on the non-addressed sites $j=1,3$ is given by $\mathrm{v}_{k,\rm homo}=\mathrm{v}_{\rm homo}=0.5$. In both PQK and classical methods, a grid-search over $C=(0.01, 0.1, 1, 10, 100, 1000)$ and $\epsilon=(0.0001, 0.001, 0.01, 0.1)$ is performed. Regarding the RBF method, results are obtained with an additional search over $\gamma=(0.1, 1, 10, (N_f\cdot\sigma_{\rm avg}^2)^{-1}, N_f^{-1})$, $N_f$ being the number of features of input data and $\sigma_{\rm avg}^2$ their average standard deviation. The number of samples is $N_{\rm samples}=40$ of which $N_{\rm train}=20$ and $N_{\rm hidden}=20$. We use $20$ different sets of data each composed of $40$ samples, from which we compute average value and standard deviation of the error. Panel (b): difference between the non-addressed sites $j=1,3$ magnetizations as a function of the measurement time for different values of the interaction strength (the different colors correspond to the different nearest neighbor interaction values $V_{\rm NN}$ shown in the legend at the top of the panel (a)). We consider a single sample in which the detuning in the addressed sites $j=0,2$ is set to the extreme cases $\mathrm{v}_0 = 1$, $\mathrm{v}_2 = 0$, while on the non addresses sites it is set to $\mathrm{v}_1 = \mathrm{v}_3 = \mathrm{v}_{\rm homo}=0.5$. The Hamiltonian parameters are the same as those of panel (a). Panel (c): density distribution of a specific potential configuration $\boldsymbol{\mathrm{v}}_k$.  On the top we show the full density distributions predicted by a PQK method at different measurement times, they are compared with the expected KS density (original~\eqref{eq:expansion} and approximated~\eqref{eq:approx}). On the bottom, we report the difference between approximated KS density and different ML predicted densities (Eq.~\eqref{eq:delta}) at different measurement times. For both top and bottom plots, the Hamiltonian paramters are $\Omega_{\rm glob}=5$ rad/$\mu$s, $\Delta_{\rm glob}=0$, $\Delta_{\rm loc}=-3.5$ rad/$\mu$s, $V_{\rm NN}=4$ rad/$\mu$s.}
\label{fig:fig2}
\end{figure*}
The original input data are mapped into a larger feature space (quantum embedding) through the dynamics of a given quantum reservoir, and the regression is then performed with the embedded data~\cite{kornjaca2024large}. As for the quantum reservoir, we consider a system of Rydberg atoms, which constitutes a highly controllable quantum platform and it has been proven to be promising in this context~\cite{bravo2022quantum,kornjaca2024large, settino2024memory}. Our protocol works as follows: to any external potential $V_k(x)$ we associate the input vector $\bm{\mathrm{v}}_{k}$ with the resulting KS density distribution $n_k(x)$ codified by the target variables $\{ u_k^{(\ell)} \}_{\ell}$. The vector $\bm{\mathrm{v}}_{k}$ is used to encode the information about the potential into the Rydberg Hamiltonian~\cite{browaeys2020many, bernien2017probing}
\begin{equation}
    \mathcal{H}_k^{\rm (Res)}=\dfrac{\Omega}{2}\sum_{j=0}^{L-1} \sigma_j^x-\sum_{j=0}^{L-1}\Delta_{k,j}n_j + \sum_{i<j=0}^{L-1} V_{ij}n_i n_j,
    \label{eq:H_res}
\end{equation}
where $L$ is the number of qubits, $\sigma_j^x = \ket{r_j}\bra{g_j}+\ket{g_j}\bra{r_j}$, $n_j = \ket{r_j}\bra{r_j}$, with $\ket{g_j}(\ket{r_j})$ being the ground (Rydberg) state of the atom $j$. Hereafter we set $\hbar=1$ and so the Hamiltonian parameters are in rad/s units. $V_{ij}=C_6/d_{ij}^6$ is the Rydberg-Rydberg interaction, depending on the inter-atomic distance $d_{ij}$ and on the interaction coefficient $C_6$, which we take as $C_6=865723.02\: \textrm{rad}\cdot\mu\textrm{m}^6\cdot\mu \textrm{s}^{-1}$~\cite{silvrio2022pulser}. $\Omega_{\rm glob}$ is the Rabi frequency of the field that couples ground and Rydberg states, and $\Delta_{k,j}$ is its associated detuning that depends on the site $j$ and on the specific configuration $k$ considered. Thus, the potential features are encoded in the local part of the detuning following the local encoding~\cite{kornjaca2024large}. 
Numerical simulations have been performed using Pulser~\cite{silvrio2022pulser}, a software package developed by Pasqal, 
which allows to simulate their architectures. 
In the following we simulate a system of four qubits.

The reservoir is prepared in the state $\ket{\psi(0)}=\ket{gggg}$, i.e. all the atoms in the ground state, and then it evolves with the Hamiltonian Eq.~\eqref{eq:H_res} up to a time $t^*$ at which we measure the following single particle mean values and the two-body correlation functions:
\begin{equation}
\bm{\mathrm{m}}_{k}=\{ \braket{\sigma_j^z}_k \}_j \cup \{ \braket{\sigma_{i}^z\sigma_{j}^z}_k \}_{i<j},  
\label{eq:measurement}
\end{equation}
where $\braket{\ldots}_k=\braket{\psi_k(t^*)|\ldots|\psi_k(t^*)}$. The new vector $\bm{\mathrm{m}}_{k}$ has size $L\left(\frac{L+1}{2} \right)$. Since the local encoding allows for the use of a number of features up to $L$, the quantum dynamics of the reservoir results in the enlargement of the feature space: 
\begin{equation}
\bm{\mathrm{v}}_{k} \in \mathbb{R}^{N_f\leq L} \longrightarrow \bm{\mathrm{m}}_{k}=\phi(\bm{\mathrm{v}}_{k}) \in \mathbb{R}^{L\left(\frac{L+1}{2} \right)},
\end{equation}
where $N_f$ is the number of features of the external potential. Once the mapping is complete, we can define the PQK $K_{\rm PQK}(\bm{\mathrm{v}}_{k},\bm{\mathrm{v}}_{l})=\bm{\mathrm{m}}_{k}\cdot\bm{\mathrm{m}}_{l}$~\cite{kornjaca2024large}. The training is numerically performed by using the Python package scikit-learn~\cite{scikit2011pedregosa}.

Once the SVR is trained, we test its performance on the hidden samples by extracting the predicted coefficients $\{ \tilde{u}_k^{(\ell), \rm ML} \}_{\ell}$ through which we can compute the predicted density distribution
\begin{equation}
    n_k^{\rm ML}(x)=\sum_{\ell=1}^{N_{\rm trunc}}\tilde{u}_k^{(\ell), \rm ML}\Psi^{(\ell)}(x).
\end{equation}
To estimate the accuracy of the prediction we define the error:
\begin{equation}
    \mathfrak{E}=\dfrac{1}{N_{\rm hidden}}\sum_{k=1}^{N_{\rm hidden}}\int dx \: \left|n_k^{\rm ML}(x) - n_k^{(\rm approx)}(x)\right|.
    \label{eq:error}
\end{equation}
The PQK results are then compared with those obtained with classical linear and RBF methods. The full procedure is schematically represented in Fig.~\ref{fig:sketch}.

\section{Results}
\label{sec:results}
In order to test our approach, we consider two  different 1D interacting fermion systems. The first scenario belongs to the so-called one-dimensional quantum chemistry. We address the problem of finding the electronic densities of a bi-atomic one-dimensional molecule. The external potential for the electrons is generated by two nuclei each with one proton, $Z=1$ and it is approximated as an attractive softened Coulomb potential, mimicking the $H_2$ molecule Hamiltonian~\cite{wagner2012reference}, see Appendix~\ref{app:one_dim}. The potential is codified by $2$ features which are the positions of the nuclei. This class of potentials has been considered in Refs.~\cite{li2016pure,snyder2013orbital,li2015understanding} as an application of classical ML techniques to the prediction of density profiles and their functionals.
The second scenario we consider is that of a triple-well potential in which the two barriers separating the wells have variable height and width, see Appendix~\ref{app:one_dim}; thus, the potential is codified by $4$ features. 1D quantum well problems have recently been addressed by using ML techniques in Ref.~\cite{sarkar2025physics} to predict the ground state wave-function structure.

For both of the models, we used a quantum reservoir of of $L=4$ interacting qubits, placed in a square geometry. The evolution of the system is driven by the Rydberg Hamiltonian~\eqref{eq:H_res}. We follow the labeling of the qubits shown in Fig.~\ref{fig:sketch}. We choose the following encoding for the classical information: 
\begin{equation}
\Delta_{k,j}=\Delta_{\rm glob}+\textrm{v}_{k,j}\Delta_{\rm loc},   
\end{equation}
where $\Delta_{\rm glob}$ is the detuning component common to all qubits and $\Delta_{\rm loc}$ is the amplitude of local detuning, namely local encoding~\cite{kornjaca2024large}. 

In the case of $H_2$, the two potential features are encoded in the detuning of two nearest-neighbor qubits, while the other two have homogeneous detuning. The features of the potential $\delta X_A$ and $\delta X_B$, carrying the information on nuclei positions (see Appendix~\ref{app:one_dim}), are first rescaled to be in the interval $[0,1]$ and then encoded in the two nearest-neighbor qubits $j=0$ and $j=2$. The detuning of the remaining qubits is homogeneous and fixed to $\Delta_{k,j=1,3}=\Delta_{\rm glob}+\rm{v}_{k,\rm homo}\Delta_{\rm loc}$. 

In the case of the triple well potential, the features $\left(h_{1,k},h_{2,k},\delta_{2,k},\delta_{3,k}\right)$, heights and widths of the two barriers (see Appendix~\ref{app:one_dim}), are first rescaled to the interval $[0,1]$ and then encoded in the Hamiltonian of the reservoir via the mapping $\left(\textrm{v}_{k,0},\textrm{v}_{k,1},\textrm{v}_{k,2},\textrm{v}_{k,3}\right)=\left(h_{1,k},h_{2,k},\delta_{2,k},\delta_{3,k}\right)$. Conversely, for the two classical methods, the rescaled features are directly used as input values for training. 

The target variables for the training procedure are  the coefficients $\{ u_k^{(\ell)}\}_{\ell}$. Their values strongly depend on the choice of the basis $\{\Psi^{(\ell)}(x)\}_{\ell}$ used to expand the density. We observe that, in both models, the space is divided into three regions, either by the two nuclei for $H_2$ or by the two barriers for the triple well. We found that an efficient choice, with some slight differences between the two cases, is to combine the three bases obtained by solving single-particle problem in the three regions: $\{\Psi_L^{(\ell)}(x)\}_{\ell=1}^{N_L}$, $\{\Psi_C^{(\ell)}(x)\}_{\ell=1}^{N_C}$ and $\{\Psi_R^{(\ell)}(x)\}_{\ell=1}^{N_R}$ with $N_L + N_R + N_C  = N_{\rm trunc}$ (more details on their construction are given in Appendix~\ref{app:building}).

\subsection{1D $H_2$ molecule}
\label{subsec:H2}
We first consider a 1D $H_2$ molecule, which is part of the class of one-dimensional chemistry models that are considered promising theoretical laboratories for real molecules~\cite{wagner2012reference,stoudenmire2012one,loos2015chemistry, ball2015chem1D, baker2015one, li2021exact, li2016pure,li2015understanding, snyder2013orbital, gedeon2021machine}. The parameters of the model are set to the values reported in Appendix~\ref{app:one_dim}. The truncated density distribution is built by using $N_L=N_C=N_R=10$, implying $N_{\rm trunc}=30$.

In Fig.~\ref{fig:fig2}(a), we report the behavior of the error of the PQK method as function of the time at which the measurement occurs. We checked different values of the nearest neighbor interaction $V_{\rm NN}=C_6/a^6$, $a$ being the lattice spacing between the atoms. The error is compared with the one of the classical linear and RBF methods. We set $\Delta_{\rm glob}=0$ for all of the qubits, while the detuning of the non-addressed ones, i.e. $j=1,3$, is set by putting $\textrm{v}_{k,\rm homo}=\textrm{v}_{\rm homo}=0.5$. 
The measurement times are rescaled in units of the frequency $\Omega_{\rm max}=\sqrt{\Omega_{\rm glob}^2 + \Delta_{\rm loc}^2}$, which is the precession frequency of the Bloch vector in the absence of interaction and global detuning.
\begin{figure}[t!]
\centering
\includegraphics[width=0.9\linewidth]{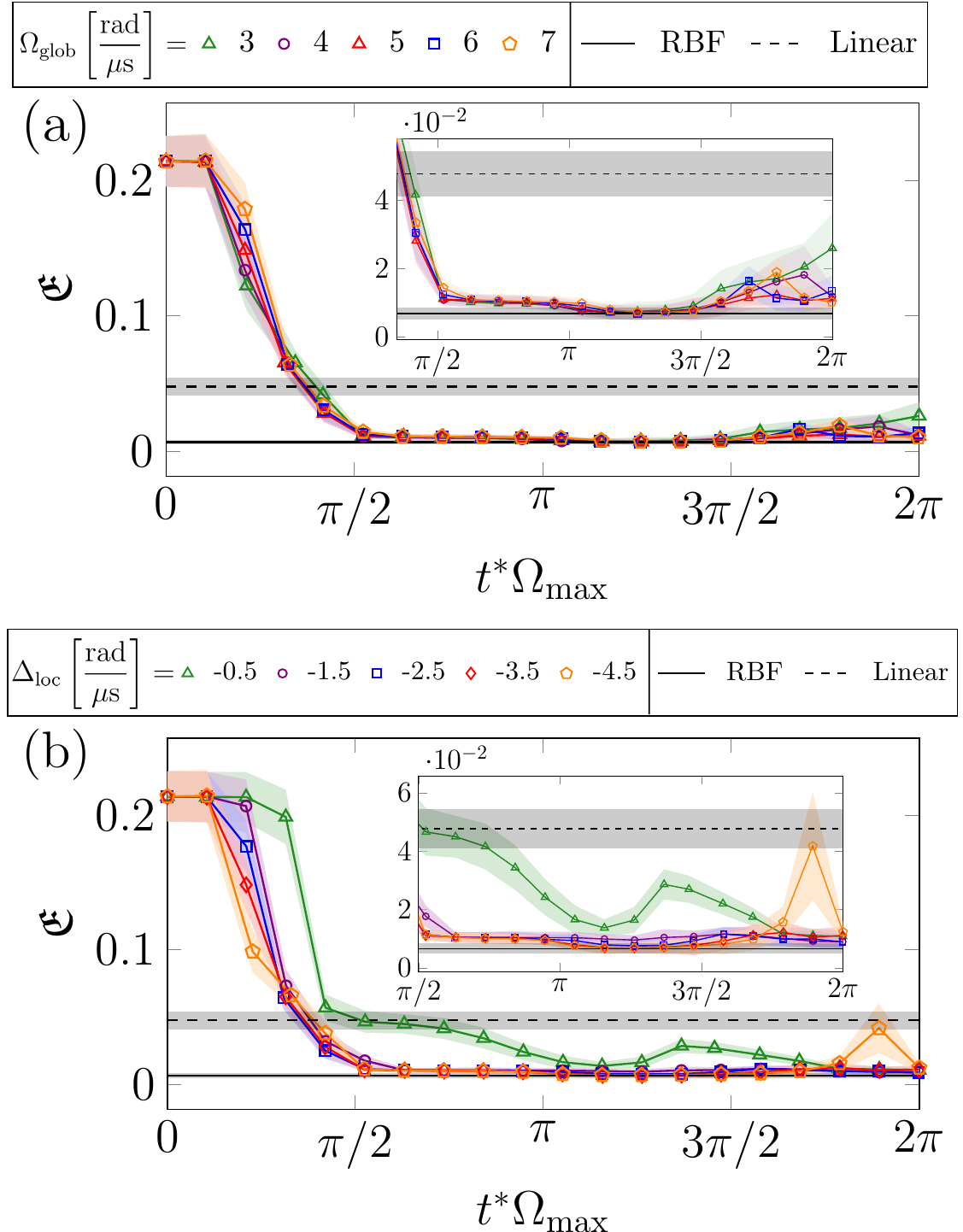}
\caption{Error of PQK SVR as a function of the measurement time for different values of the global Rabi frequency $\Omega_{\rm glob}$ [panel (a)] and for different values of the local detuning amplitude $\Delta_{\rm loc}$ [panel (b)]. In both cases the error is compared with those obtained with classical linear and RBF kernel methods. The Hamiltonian parameters of panel (a) are: $\Delta_{\rm glob}=0$, $\Delta_{\rm loc}=-3.5$ rad/$\mu$s, $V_{\rm NN}=4$ rad/$\mu$s, $\textrm{v}_{\rm homo}=0.5$. The Hamiltonian parmeters of panel (b) are: $\Delta_{\rm glob}=0$, $\Omega_{\rm glob}=5$ rad/$\mu$s, $V_{\rm NN}=4$ rad/$\mu$s, $\textrm{v}_{\rm homo}=0.5$. Grid search and computation of the error are performed as in Fig.~\ref{fig:fig2}(a).}
\label{fig:fig3}
\end{figure}
The behavior of the error is characterized by the presence of a small plateau at short times followed by a sudden drop before stabilizing. After the drop, the error is noticeably smaller than the one obtained with a classical linear kernel method. Once stabilized, for this specific choice of the Hamiltonian parameters, the error has a value that is comparable with the one obtained with a RBF kernel method.  Overall, the best performance of the PQK method is obtained in the interval $\pi \lesssim t^*\Omega_{\rm max} \lesssim 3\pi/2$. The stabilization time depends on the reservoir interaction strength $V_{\rm NN}$ and increases with the decrease in the interaction due to the chosen encoding. Indeed, the encoding is performed only on two qubits, therefore the non-addressed qubits $j=1,3$ have the same magnetization dynamics $\braket{\sigma_1^z}=\braket{\sigma_3^z}$ in the non-interacting limit; moreover, the correlations $\braket{\sigma_0^z\sigma_1^z}$ and $\braket{\sigma_1^z\sigma_2^z}$ respectively coincide with $\braket{\sigma_0^z\sigma_3^z}$ and $\braket{\sigma_1^z\sigma_3^z}$. Thus, the number of features is effectively $N_f=7$, instead of being $N_f=L\left(\frac{L+1}{2}\right)=10$. The van der Waals $1/d_{ij}^6$ interaction between addressed and non-addressed sites induces inhomogeneity in the dynamics of the observables, bringing the feature space dimension back to $N_f=10$. The inhomogeneity is enhanced for stronger interactions, and it occurs at shorter times, implying the error drops faster for larger interactions. This implies that the role of the interactions is important in our protocol, not only to keep the dynamics of the system within the coherence time of current NISQ devices, but it is actually beneficial for the overall performance.

Fig.~\ref{fig:fig2}(b) gives a visual description of the latter observation. We define the local magnetizations $M_j^z = \braket{\sigma_j^z}$ ($j=0,1,2,3$) and report the dynamics of the difference between those of the non-addressed sites $j=1,3$ in a reservoir with $\textrm{v}_{\rm homo}=0.5$, $\textrm{v}_0=1$ and $\textrm{v}_2=0$. We are basically considering a specific sample in which the addresses sites have the two extreme possible values of detuning. At short times, the magnetization difference is quite small and almost independent on the interaction strength. At times $t^*\Omega_{\rm max}\approx \pi/4$, which correspond to the drop of the error, the magnetization difference becomes noticeably different from zero and bigger for larger interactions. Therefore, the distinguishability between observables depends on the strength of the reservoir interaction, making the error in this type of encoding sensitive to it.

\begin{figure}[t!]
\centering
\includegraphics[width=0.9\linewidth]{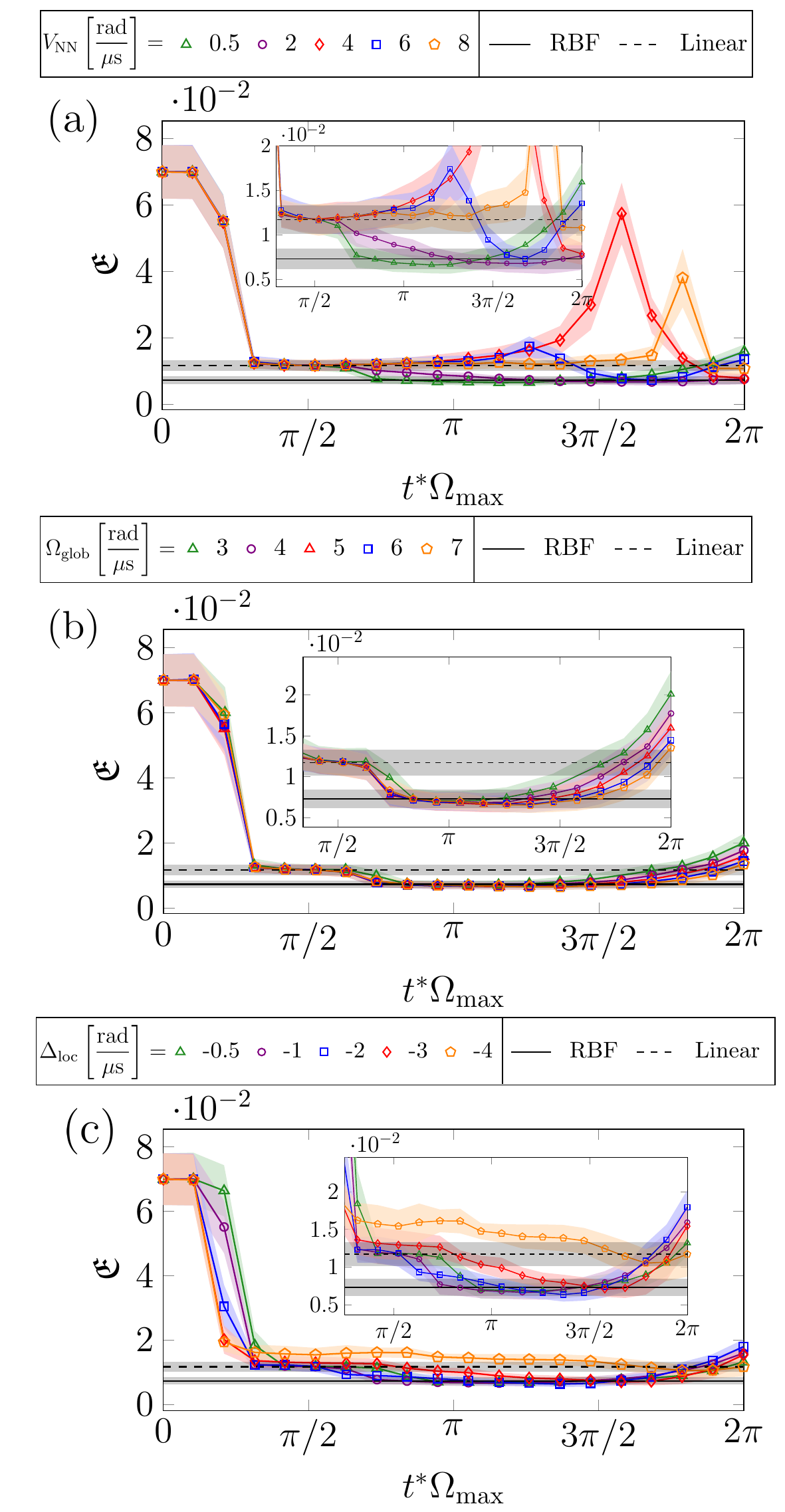}
\caption{Error of the PQK method as a function of the measurement time for different values of the Hamiltonian parameters $V_{\rm NN}$ [panel (a)], $\Omega_{\rm glob}$ [panel (b)] and $\Delta_{\rm loc}$ [panel (c)]. In any case the global detuning term is set to $\Delta_{\rm glob}=5$ rad/$\mu$s, while the other parameters are set to $\Omega_{\rm glob}=5$ rad/$\mu$s, $\Delta_{\rm loc}=-1$ rad/$\mu$s [panel (a)], $V_{\rm NN}=0.5$ rad/$\mu$s, $\Delta_{\rm loc}=-1$ rad/$\mu$s [panel (b)] and $\Omega_{\rm glob}=5$ rad/$\mu$s, $V_{\rm NN}=0.5$ rad/$\mu$s [panel (c)]. In any case, results are compared with classical RBF and Linear methods, grid search and computation of the error are performed as in Fig.~\ref{fig:fig2}(a).}
\label{fig:fig4}
\end{figure}
In Fig.~\ref{fig:fig2}(c), we report the predicted density for a given hidden sample $k$ obtained with different methods,
and we compare it with the expected one.
The top panels show a comparison of the KS density profile $n_k$ and its truncated version $n_k^{(\rm approx)}$ with the profiles predicted through the PQK method. We extract the density profile at three different measurement times: from left to right $t^*\Omega_{\rm max}=\pi/10, \pi/4, 5\pi/4$. For $t^*\Omega_{\rm max}=\pi/10$, the error is in the initial plateau and the predicted density has a poor agreement with the KS density profile. For $t^*\Omega_{\rm max}=\pi/4$, we are in the region where the error drops, but it is still not stabilized. The agreement between the predicted and target density improves but it is still not optimal. $t^*\Omega_{\rm max}=5\pi/4$ belongs to the time interval $[\pi,3\pi/2]$ in which the PQK shows overall the best performance. We also mention that there is  good agreement between $n_k$ and $n_k^{(\rm approx)}$, proving that $N_{\rm trunc}=30$ is sufficient to accurately reconstruct the density profile.

The bottom panels of Fig.~\ref{fig:fig2}(c) report the quantity
\begin{equation}
    \delta n_k^{\rm ML}(x)=\left|n_k^{(\rm approx)}(x)-n_k^{\rm ML}(x) \right|,
    \label{eq:delta}
\end{equation}
where $n^{\rm ML}_k(x)$ is the ML-predicted density through SVR methods: $\rm ML = \rm lin, \: RBF, \: PQK$. We observe that the RBF-predicted density approximates well the expected KS density in the entire space domain $x$. On the contrary, the prediction of the linear method fails, with a dominant error in the center of the spatial domain, in the area where the nuclei are located. Regarding the PQK method, as expected, at increasing measurement times the prediction works better, with $\delta n^{\rm PQK}$ that is comparable with $\delta n^{\rm RBF}$ at $t^*\Omega_{\rm max}=5\pi/4$.

We investigated the performance of the PQK method on the values of the other Hamiltonian parameters $\Omega_{\rm glob}$ and $\Delta_{\rm loc}$. We fix the nearest-neighbor interaction to $V_{\rm NN}=4$ rad/$\mu$s and analyze the error behavior for different values of the single-atom Hamiltonian parameters. Fig.~\ref{fig:fig3}(a) reports the dependence of the error on the measurement time for different values of the global Rabi frequency. 
We see that up to measurement times $t^*\Omega_{\rm max}\gtrsim 3\pi/2 $ the overall behavior of the error shows a perfect scaling with $\Omega_{\rm glob}$. Only in the dropping region we observe a mild dependence on the the Rabi frequency, which affects the slope at which the error drops. This is due to the fact that, for our choices of parameters, the global Rabi frequency is the dominant energy scale for the single atom dynamics. This implies that it dictates the time scale for the mixing of the information within the single atom subspace, which is then redistributed to the rest of the system via the interaction term.
It is interesting to notice that for large measurement times, the error starts growing with an irregular dependence on $\Omega_{\rm glob}$.

On the other hand, the amplitude of local detuning affects more significantly the error. Fig.~\ref{fig:fig3}(b) shows its behavior as a function of measurement time for different values of $\Delta_{\rm loc}$. 
The worst performance is obtained for a small value of the detuning amplitude ($\Delta_{\rm loc}=-0.5$ rad/$\mu$s), with the error that drops but does not stabilize. For the other values of $\Delta_{\rm loc}$, we observe behaviors which are alike and overall good performances, competitive with the RBF kernel method in the time interval $[\pi/\Omega_{\rm max},3\pi/(2\Omega_{\rm max})]$. 
\subsection{Triple-well potential}
\label{subsec:Triple}
We now analyze the performance of the PQK method for predicting the fermionic density of the triple-well model, whose parameter values are reported in Appendix~\ref{app:one_dim}. The truncated density distribution is built by using $N_L=N_C=N_R=6$, corresponding to $N_{\rm trunc}=18$. In this case, the potential is codified by $4$ features which are encoded in the detunings of all the qubits of the reservoir Hamiltonian. In Fig.~\ref{fig:fig4} we report the comparison between the error of the PQK method and those of classical linear and RBF kernel methods. In this case, we fix $\Delta_{\rm glob}=5$ rad/$\mu$s and so, with negative $\Delta_{\rm loc}$, the rescaling maximum frequency is $\Omega_{\rm max}=\sqrt{\Omega_{\rm glob}^2 + \Delta_{\rm glob}^2}$. We observe that error of the classical linear method is much smaller in this case than in the $H_2$ case, indicating that data are now well fitted by a hyperplane. The results are improved further by using a RBF method, for which the error is comparable with those of $H_2$. Regarding the PQK method, the error possesses the same general features observed in the $H_2$ case: at short measurement times there is a short plateau, then a sudden drop followed by a stabilization whose width depends on Hamiltonian parameters. 

In Fig.~\ref{fig:fig4}(a), we analyze the effect of the reservoir nearest-neighbor interaction. We observe that the error weakly depends on the latter at short measurement times. At $t^*\Omega_{\rm max}\gtrsim\pi /2$, a clear and irregular dependence on the reservoir interaction appears. In general, we note that, in the measurement time interval considered, the PQK method performs better for weak reservoir interactions. For any value of the reservoir interaction, it is possible to identify measurement time windows in which the PQK method performs better than the linear kernel. 
It is also possible to identify time windows in which the PQK method competes with the RBF kernel, except for $V_{\rm NN}=8$ rad/$\mu$s. In general, for larger interactions, the error behavior is less stable, with the appearance of sudden peaks. Fig.~\ref{fig:fig4}(b) reports the error behavior for different values of the Rabi frequency. We observe that during the drop of the error and its stabilization, $\Omega_{\rm glob}$ weakly affects the performances of the method. After stabilization, the error grows showing an inverse relationship with the Rabi frequency. Finally, we comment on the effect of the local detuning amplitude $\Delta_{\rm loc}$, see Fig.~\ref{fig:fig4}(c). Similarly to the $H_2$ case, the error drops faster for larger $\Delta_{\rm loc}$. However, after the drop, larger $\Delta_{\rm loc}$ show the worst overall performance.  

In Appendix~\ref{app:additional}, we also analyzed additional features of the error like its scaling with the number of training samples. We observe that, for a specific set of reservoir parameters and varying the number of training samples, the PQK method remains competitive with the RBF kernel and better than the linear kernel.

\section{Conclusions and outlook}
\label{sec:conclusions}
We proposed a PQK SVR method to predict the density profile of 1D continuous fermionic systems. Training and test data are obtained by a DFT approach, solving two paradigmatic models: the 1D $H_2$ molecule and a system of two fermions in a triple-well potential. The method that we used is based on processing input data through a few-qubits quantum reservoir implementable in Rydberg atom architectures. Input data are encoded into the reservoir, then, at a specific measurement time, a set of observables is measured; it contains the processed data that are used to construct the kernel matrix. The performance of the method is quantified by the error between the expected output densities of the hidden test data and their ML prediction, which is then compared with that of classical kernel methods such as RBF and linear.

The reservoir parameters are essential to obtain the best performances of the method. We found that among all, the measurement time plays the most important role. In fact, the reservoir is initialized in a polarized state in which all the atoms are in the ground state and data are completely undistinguishable at the beginning. By increasing the measurement time, the distinguishability between the processed data increases, and, thus, the method is expected to perform better. We found that the error has a short plateau at small measurement times before dropping, indicating that the reservoir needs a minimum of time to process the input information. This is true regardless of the problem at hand and/or of the choice of parameters, indicating that it is linked to the microscopic dynamics of the Rydberg atoms, which needs to spread the information across the system.
Following this time, the error drops suddenly and significantly, signaling the onset of the mixing. Eventually, the error stabilizes at values which are comparable with those of classical methods. Specifically, for both $H_2$ and triple-well, the PQK method often outperforms the linear kernel results and can be competitive with the RBF kernel. The order of magnitude of the times at which the error stabilizes are a few tenths of $\mu$s. 

We observe that this general error behavior is common to both models analyzed for different choices of Hamiltonian parameters. In particular, in both cases, different choices of the Rabi frequency weakly affect the method performance [see Fig.~\ref{fig:fig3}(a) and Fig.~\ref{fig:fig4}(b)]; on the other hand, weak or large reservoir interactions and local detuning amplitudes can be detrimental to the method performance by altering the aforementioned behavior. 

We remark that our method requires only a reservoir composed of $4$ interacting qubits to predict faithfully the density structure of the 1D fermionic systems of interest, see Fig.~\ref{fig:fig2}(c). The crucial point is to choose the best measurement time that, for a particular set of the reservoir Hamiltonian parameters, minimizes the error. The reservoir main role is to define a complex mapping $\bm{\mathrm{m}}=\phi(\bm{\mathrm{v}})$ that enlarges the feature space and it is highly controllable by its parameters $t^*,\Omega_{\rm glob}, \Delta_{\rm glob},\Delta_{\rm loc},V_{\rm NN}$. The complex mapping is defined by the dynamics of a controllable quantum system and thus its performance can be also experimentally tested on real quantum simulators. In this regard, the implementation of the method in large quantum simulators allows for the realization of a controllable mapping in an enlarged feature space that cannot be easily replicated on a classical hardware. In this work, we considered a Rydberg atom Hamiltonian, but in general the same approach can be tested by using Hamiltonians implementable in different plaftorms including trapped ions~\cite{blatt2012quantum}, ultra-cold atoms in optical lattices~\cite{bloch2012quantum}, atoms in cavities~\cite{bentsen2019treelike} etc.

We mention that we studied the method on two specific 1D problems. The verification of its validity can serve as a starting point for its application in 1D models with different potential landscapes and real 3D molecules~\cite{brockherde2017bypassing,bogojeski2020quantum,beaulieu2024robust,chuiko2025predicting, Lomonaco2024quantum, patil2025machine}. Moreover, starting from these results, the PQK method can be tested on the step-forward of the DFT, which is evaluating the functionals from the predicted density distributions with a second PQK SVR machine~\cite{bogojeski2020quantum, li2016pure, snyder2013orbital}. Different classes of PQK~\cite{huang2021power,schnabel2024quantum} can be used as well.

\section{Acknowledgments}
We thank A.  Palamara and J.  Settino for fruitful discussions. This research was partially supported by the PNRR MUR project PE0000023-NQSTI through the secondary projects ``QuCADD" and ``ThAnQ", and the Italian Ministry of Health with the project “CAL.HUB.RIA” - CALABRIA HUB PER RICERCA INNOVATIVA ED AVANZATA. Code: T4-AN-09, CUP: F63C22000530001. Numerical Simulations of the KS calculations were run on the Galileo100 machine of CINECA with the project IsCc5$\_$DIAGME.

\appendix

\section{1D fermionic models}
\label{app:one_dim}
\begin{figure}[htbp]
\centering
\includegraphics[width=\linewidth]{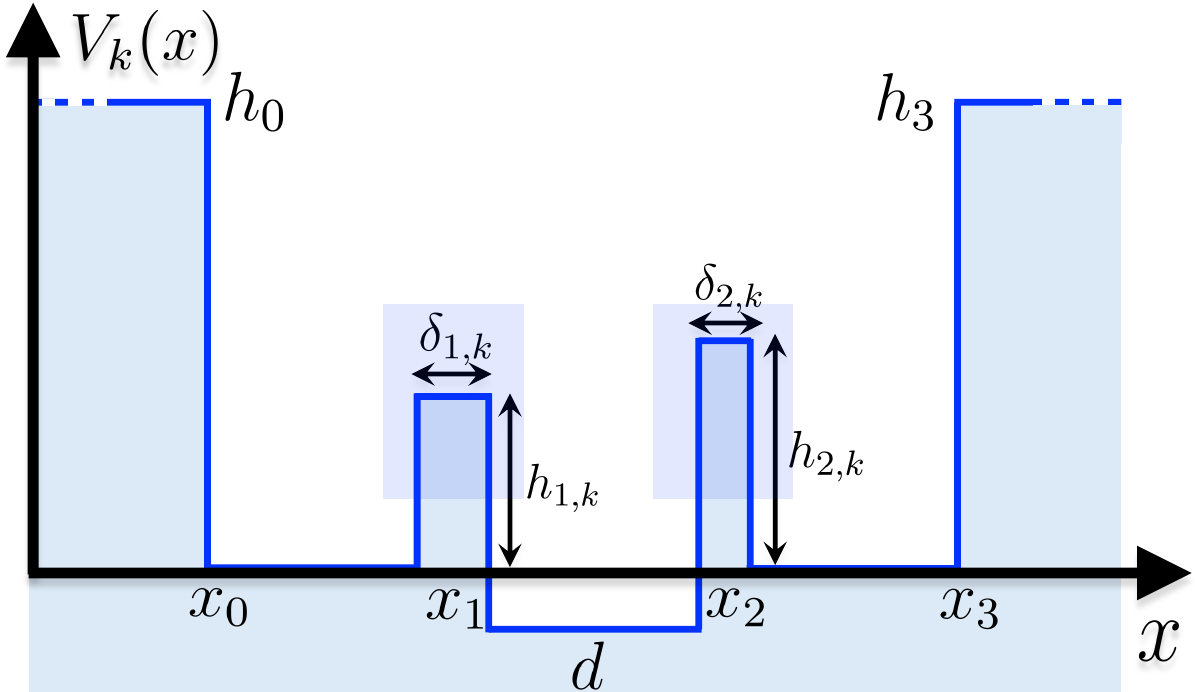}
\caption{Sketch of the triple-well potential defined in Eq.~\eqref{eq:potential}.}
\label{fig:potential_sketch}
\end{figure}
% 
%specify models and parameters
We consider two different 1D fermionic models both consisting of Hamiltonians of the type~\eqref{eq:H_ferm}
\begin{equation}
\mathcal{H}^{(2\rm F)}_k = \sum_{i=1,2}\left[T_i + V_k(x_i)\right] + U(|x_1 - x_2|),  
\end{equation}
with
\begin{equation}
T_i =-\dfrac{1}{2m_i}\dfrac{\partial^2}{\partial x_i^2},
\quad
U(|x_1-x_2|)=\dfrac{1}{|x_1-x_2|+1},
\end{equation}
with $\hbar=1$. $m_1$ and $m_2$ are the masses of the fermions, and $U$ describes their softened Coulomb repulsive interaction. The external potential $V_k(x)$ defines the problem.

We first consider the $H_2$ one-dimensional diatomic molecule. The potential is given by two hydrogen nuclei ($Z=1$) at fixed positions $X_A, X_B$ that interact with the electrons through an attractive softened Coulomb interaction
\begin{equation}
    V_k(x)=-\dfrac{1}{|x-X_{A,k}|+1} - \dfrac{1}{|x-X_{B,k}|+1},
    \label{eq:H2_pot}
\end{equation}
where the index $k$ labels different potential configurations, given here by the different positions of the nuclei. We model the nuclei positions as $X_{A,k}=X_{A}^0 + \delta X_{A,k}$ and $X_{B,k}=X_{B}^0 + \delta X_{B,k}$, where $X_{A}^0$ and $X_B^0$ are kept fixed, while $\delta X_{A,k}$ and $\delta X_{B,k}$ are varied in a specific interval. $\delta X_{A,k}$ and $\delta X_{B,k}$ are the potential features.
We impose Open Boundary Conditions (OBC) and consider the spatial domain to be $x\in[0,L_x]$, with $L_x=10$ a.u.; the potential configurations are generated by fixing $X_{A}^0=4$ a.u.,  $X_B^0=6$ a.u. and by extracting randomly the two potential features $\delta X_A$ and $\delta X_B$ in uniform distributions with range $[-0.5,0.5]$ a.u.. The masses of the fermions are fixed to $m_1 = m_2 = 1$ a.u..

The second potential of interest is the triple-well potential given by 

\begin{equation}
    V_k(x)=
    \begin{cases}
        h_0 \: \: \textrm{if} \: \: x \leq x_0, \\
        0 \: \: \textrm{if} \: \:  x_0 <x\leq x_1 - \delta_{1,k}/2, \\
        h_{1,k}\: \: \textrm{if} \: \:  x_1 - \delta_{1,k}/2 <x\leq x_1 + \delta_{1,k}/2, \\
        d \: \: \textrm{if} \: \:  x_1 + \delta_{1,k}/2 <x\leq x_2 - \delta_{2,k}/2, \\
        h_{2,k} \: \: \textrm{if} \: \:  x_2 - \delta_{2,k}/2 <x\leq x_2 + \delta_{2,k}/2, \\
        0 \: \: \textrm{if} \: \:  x_2 + \delta_{2,k}/2 <x\leq x_3, \\
        h_3 \: \: \textrm{if} \: \: x > x_3, \\
    \end{cases}
    \label{eq:potential}
\end{equation}    
which is sketched in Fig.~\ref{fig:potential_sketch}. In this case, the potential features are the heights and widths of the two central barriers $h_{1,k}, h_{2,k}, \delta_{1,k}, \delta_{2,k}$. The spatial domain is $x\in [0,L_x]$, with $L_x=10$ a.u., the masses of the fermions are set to $m_1=m_2=0.5$ a.u. and Periodic Boundary Conditions (PBC) are imposed. The potential parameters are set to $x_0=0.5$ a.u., $x_1=3.5$ a.u., $x_2=6.5$ a.u., $x_3=9.5$ a.u., $h_0=h_3=5$ a.u. and $d=-0.2$ a.u.; the heights and widths of the barriers are extracted randomly from uniform distributions with the ranges $h_{1,k}, h_{2,k} \in [0.8,1.6]$ a.u. and $\delta_{1,k}, \delta_{2,k} \in [0.2,0.5]$ a.u.. In this case, data are prepared by using the Julia package QuantumOptics.jl~\cite{kramer2018quantum}.

\section{Kohn-Sham equation}
\label{app:KS}
Let us consider a system of $N$ interacting fermions described by
\begin{equation}
\mathcal{H}^{(N\rm F)} = \sum_{i=0}^{N-1}\left[T_i + V(\boldsymbol{r}_i)\right] + \sum_{i<j=0}^{N-1}U(|\boldsymbol{r}_i-\boldsymbol{r}_j|),   
\end{equation}
where the interaction $U$ is a repulsive Coulomb interaction and $N$ is the number of fermions. The exact resolution of the interacting system is a non-trivial problem that requires the computation of a complex many-body wave function. The complexity of the problem can be circumvented by recurring to approximated methods that often well capture the structure of the system. One of these is Density Functional Theory (DFT)~\cite{kohn1965self, hohenberg1964inhomogeneous, kohn1999nobel, giuliani2005quantum}, an approximate modeling method whose basic idea is to express the interacting problem in terms of the fermionic density distribution, without accessing the exact many-body wave function of the system. In fact, the first Hohenberg-Kohn theorem states that any ground state property of the system can be expressed as a functional of the density~\cite{hohenberg1964inhomogeneous}. Moreover, following the treatment of Kohn-Sham, it is possible to find the density distribution through the resolution of an effective non-interacting self-consistent equation, namely the Kohn-Sham (KS) equation~\cite{kohn1965self}:
\begin{equation}
\left[ T + V(\boldsymbol{r}) + V_{\rm H}(\boldsymbol{r}) + V_{\rm xc} (\boldsymbol{r})\right]\varphi_i(\boldsymbol{r})=E_i\varphi_i(\boldsymbol{r}),
\end{equation}
where the eigenstates $\varphi_i$ are called KS orbitals, $V_{\rm H}$ is the Hartree potential that accounts for repulsive Coulomb interactions, and $V_{\rm xc}$ is the exchange-correlation potential that accounts for exchange and quantum correlations effects. The density profile can be expressed in terms of the KS orbitals as $n(\boldsymbol{r})=\sum_{i=0}^{N-1}|\varphi(\boldsymbol{r})|^2$. Both Hartree and exchange-correlation potentials depend on the density, thus the problem must be solved with a self-consistent approach. One begins with a guess for the KS orbitals $\{\varphi_i^{[1]}(x)\}_i$ which define the guessed density distribution $n^{[1]}(x)$. The guessed density is used to obtain the potentials and so the new set of KS orbitals $\{\varphi_i^{[2]}(x)\}_i$ and, consequently, the density $n^{[2]}(x)$. The procedure is iterated until convegence is reached and the KS density distribution is obtained.

In our specific case, we consider a 1D system composed of two fermions interacting via a softened Coulomb interaction. The KS equation is 
\begin{align}
&\left[ T +V_{\rm KS}(x)\right]\varphi_i(x)=E_i\varphi_i(x),\\[3pt] 
&V_{\rm KS}(x) = V(x) + V_{\rm H}(x) + V_{\rm xc}(x),
\end{align}
where the the Hartree potential is
\begin{equation}
    V_{\rm H} (x)=\int dx' \dfrac{n(x')}{|x-x'|+1}
\end{equation}
and the exchange-correlation potential is given in LDA~\cite{Entwistle2016local} by
\begin{equation}
V_{\rm xc}(x)=\left( - 1.19 + 1.77 n(x) - 1.37 n(x)^2\right) n(x)^{0.604}.   
\end{equation}
Since we consider a system composed of two fermions, the density distribution is computed by
\begin{equation}
n(x)=|\varphi_0(x)|^2 +|\varphi_1(x)|^2,
\end{equation}
$\varphi_0(x)$ and $\varphi_1(x)$ being the first two levels of the KS Hamiltonian $\mathcal{H}_{\rm KS}=T + V_{\rm KS}$. 

Regarding the self-consistent resolution, we consider as initial KS orbitals those of the non-interacting system. The $s$-th iteration step is
\begin{equation}
\{\varphi_i^{[s]}(s) \}_i, \:n^{[s]}(x)\rightarrow V_{\rm KS}^{[s]}(x) \rightarrow \{\varphi_i^{[s],\rm KS}(s) \}_i,\: n^{[s], \rm KS}(x).
\end{equation}
The density $n^{[s], \rm KS}(x)$ is not yet the one that is entered into the potential at the next iteration. Indeed, to accelerate the convergence of the algorithm, we use Pulay mixing~\cite{pulay1980convergence}. To do so, we introduce the difference
\begin{equation}
    R^{[m]}(x)= n^{[m], \rm KS}(x) - n^{[s]}(x), \quad m = \overline{s},\ldots, s;
\end{equation}
where $\overline{s}=s-\textrm{cut}+1$, with $\textrm{cut}=\textrm{min}(\mu,s)$ and $\mu$ is a memory factor. We observe that $\textrm{cut}=1,\ldots,s$ and so $\overline{s}=s,\ldots,1$. For $\mu<s$, the number of $R^{[m]}$ considered corresponds to $\mu$, while for $\mu > s$, the number $R^{[m]}$ considered strictly corresponds to $s$. Given this definition, it is possible to introduce the next-iteration density profile
\begin{equation}
n^{[s+1]}(x)=\sum_{m=\overline{s}}^s c_m \left(n^{[m]}(x) + \alpha R^{[m]}(x)\right);    
\end{equation}
we first observe that the next-iteration density distribution is computed by mixing the previous iterations densities with their associated KS densities. The number of previous iterations taken into account is controlled by the memory factor $\mu$. The expansion coefficients are found by solving a set of equations involving inner products between the functions $R^{[m]}(x)$~\cite{pulay1980convergence}; the coefficient $\alpha$ is a mixing parameter that is tuned to guarantee better convergence. Once Pulay mixing is performed, the iteration step can be updated as
\begin{equation}
n^{[s]}(x)\rightarrow V_{\rm KS}^{[s]}(x) \rightarrow n^{[s], \rm KS}(x) \overset{\rm Pulay}{\rightarrow} n^{[s+1]}(x);
\end{equation}
the procedure is repeated until convergence is reached: $\int dx \: |n^{[s+1]}(x)-n^{[s]}(x)|<\varepsilon$, $\varepsilon$ being a sufficiently small parameter. The final $n^{[s]}(x)$ is our KS density profile.

\section{Building the basis}
\label{app:building}
In this section, we describe the procedure used for constructing the base $\{ \Psi^{(\ell)}(x)\}_{\ell}$ in which the density is expanded~\eqref{eq:expansion}. We first consider $H_2$ and define the potentials
\begin{align}
&V_L(x)=h_{\rm basis}\mathcal{P}(X_A,L_x),\nonumber\\
&V_C(x)=h_{\rm basis}\left(\mathcal{P}(0,X_A)+\mathcal{P}(X_B,L_x)\right),\nonumber\\
&V_R(x)=h_{\rm basis}\mathcal{P}(0,X_B),
\end{align}
where we introduced the piece-wise function $\mathcal{P}(z,y)=\theta(z-x)-\theta(y-x)$, with $z<y$ and $\theta(x)$ being the Heaviside step function. The function $\mathcal{P}(z,y)$ is one in the interval $[z,y]$ and zero elsewhere. Therefore, we defined three single-well potentials with wells in $[0,X_A]$, $[X_A,X_B]$ and $[X_B,L_x]$ for respectively $V_L(x)$, $V_C(x)$ and $V_R(x)$, with height $h_{\rm basis}$ that is fixed to $20$ a.u.. The basis states are built by diagonalizing the single-particle Hamiltonians $\mathcal{H}_L=T+V_L$, $\mathcal{H}_C=T+V_C$ and $\mathcal{H}_R=T+V_R$ to obtain the three sets of states $\{\Psi_L^{(\ell)}(x)\}_{\ell}$, $\{\Psi_C^{(\ell)}(x)\}_{\ell}$ and $\{\Psi_R^{(\ell)}(x)\}_{\ell}$; subsequently, we perform the truncation by collecting the first $N_L, \: N_C$ and $N_R$ number of states from the the three sets. Then, we construct a new set $\{ \tilde{\Psi}^{(\ell)}(x)\}_{\ell=1}^{N_{\rm trunc}}$ such that
\begin{align}
&\tilde{\Psi}^{(\ell)}(x) = \Psi_L^{(\ell)}(x), \quad \ell\in\left[1,N_L\right],\nonumber\\
&\tilde{\Psi}^{(\ell+N_L)}(x) = \Psi_C^{(\ell)}(x), \quad \ell\in\left[1,N_C\right],\nonumber\\
&\tilde{\Psi}^{(\ell+N_L+N_C)}(x) = \Psi_R^{(\ell)}(x), \quad\ell\in\left[1,N_R\right],
\label{eq:new_set}
\end{align}
where $N_{\rm trunc}=N_L + N_C + N_R$. Finally, the new set of states is orthonormalized through the Gram-Schmidt process:
\begin{equation}
\Psi^{(\ell)} = \tilde{\Psi}^{(\ell)} - \sum_{\ell'=1}^{\ell-1}\braket{\Psi^{(\ell ')},\tilde{\Psi}^{(\ell)}}\Psi^{(\ell ')},
\; \; \; \;
\Psi^{(\ell)} \rightarrow \Psi^{(\ell)} / \|\Psi^{(\ell)}\|, 
\label{eq:ortho}
\end{equation}
where $\Psi_{1}=\tilde{\Psi}_{1}$ and the scalar product is $\braket{\Psi_{\ell '},\tilde{\Psi}_{\ell}}=\int \: dx \: \Psi_{\ell '}^*(x) \tilde{\Psi}_{\ell '}^*(x) $. 

In the case of the triple-well potential problem, the single-particle Hamiltonians from which the basis is built have the following potential shapes:
\begin{align}
&V_L(x)=h_{\rm basis}\left(\mathcal{P}(0,x_0) + \mathcal{P}(x_1,L_x) \right),\nonumber\\
&V_C(x)=h_{\rm basis}\left(\mathcal{P}(0,x_1)+\mathcal{P}(x_2,L_x)\right)+d \:\mathcal{P}(x_1,x_2),\nonumber\\
&V_R(x)=h_{\rm basis}\left(\mathcal{P}(0,x_2) + \mathcal{P}(x_3,L_x) \right),  
\end{align}
the parameters used are $h_{\rm basis}=20$ a.u. and $d=-0.2$ a.u.. Once the Hamiltonians $\mathcal{H}_L$, $\mathcal{H}_C$ and $\mathcal{H}_R$ are diagonalized, we extract our basis states by using Eqs.~\eqref{eq:new_set} and~\eqref{eq:ortho}.

\section{Support Vector Regression}
\label{app:SVR}
\begin{figure}[!h]
\centering
\includegraphics[width=0.85\linewidth]{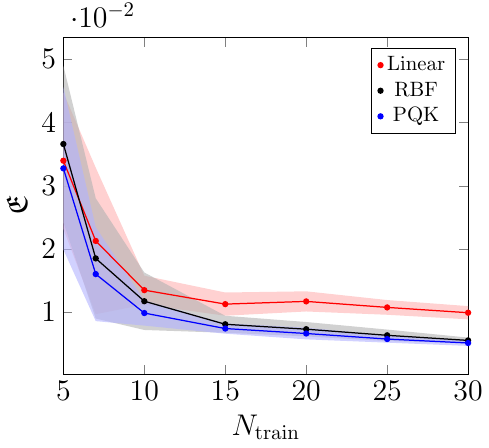}
\caption{Scaling of the error with the number of training samples $N_{\rm train}$, $N_{\rm hidden}=20$ is kept fixed. The reservoir parameters for the PQK method are $\Omega_{\rm glob}=5$ rad/$\mu$s, $\Delta_{\rm glob}=5$ rad/$\mu$s, $\Delta_{\rm loc}=-1.1$ rad/$\mu$s, $a=10\: \mu$m ($V_{\rm NN}\approx 0.866$ rad/$\mu$s), $t^* =0.6 \: \mu$s.  The grid search is performed as in Fig.~\ref{fig:fig2}(a).}
\label{fig:err_scaling}
\end{figure}
Support Vector Regression (SVR) is a supervised learning algorithm whose aim is to find the best hyperplane that fits a certain set of input data~\cite{scikit2011pedregosa, hearst1998support, kecman2005support, pml1Book}. To explain how it works, let us consider a set of data 
\begin{equation}
    \{\bm{\mathrm{x}}_k, \mathrm{y}_k \}, \quad \bm{\mathrm{x}}_k\in \mathbb{R}^{N_f}, \mathrm{y}_k \in \mathbb{R}, \quad k=1,\ldots,N_{\rm train},
\end{equation}
where $\bm{\mathrm{x}}_k$ are the input data composed each of $N_f$ number of features and $\mathrm{y}_k$ are their corresponding outputs; the goal of the SVR is to fit the data through the linear relation
\begin{equation}
    \tilde{\mathrm{y}}(\bm{\mathrm{x}})=\bm{\mathrm{w}}\cdot \bm{\mathrm{x}} + b,
    \label{eq:linear}
\end{equation}
where $\bm{\mathrm{w}} \in \mathbb{R}^{N_f}$ defines the hyperplane that fits the data and $b\in \mathbb{R}$. For doing so, the requirement is to solve a specific optimization problem under the constraint
\begin{equation}
    |\mathrm{y}_k - \tilde{\mathrm{y}}(\boldsymbol{\mathrm{x}}_k)|\leq \epsilon,
    \label{eq:tube}
\end{equation}
which leads to the solution
\begin{equation}
\tilde{\mathrm{y}}(\boldsymbol{\mathrm{x}}) = \sum_{k \in \rm SV}(\alpha_k - \alpha_k^*)\boldsymbol{\mathrm{x}}_k\cdot\boldsymbol{\mathrm{x}}+b,  
\end{equation}
where the Support Vectors (SV) are the data points that lie on the margin or outside the $\epsilon$-tube defined by Eq.~\eqref{eq:tube}, the coefficients $\alpha_k$ and $b$ are found by the resolution of the optimization problem.

This type of algorithm assumes that the input and output data are connected through a linear relation~\eqref{eq:linear}. In general, the relation can be more complex and not captured by a hyperplane whose dimensionality corresponds to the number of features. Thus, the method can be improved by mapping the data in a space of higher dimensionality by enlarging the feature space:
\begin{equation}
\bm{\mathrm{x}}_k \in \mathbb{R}^{N_f} \longrightarrow \phi(\bm{\mathrm{x}}_k) \in \mathbb{R}^{N_F>N_f}.    
\end{equation}
Once the feature space is enlarged, the SVR predicted output is
\begin{align}
\tilde{\mathrm{y}}(\boldsymbol{\mathrm{x}}) &= \sum_{k \in \rm SV}(\alpha_k - \alpha_k^*)\phi(\boldsymbol{\mathrm{x}}_k)\cdot\phi(\boldsymbol{\mathrm{x}})+b \\
&\coloneqq \sum_{k \in \rm SV}(\alpha_k - \alpha_k^*)K(\boldsymbol{\mathrm{x}}_k,\boldsymbol{\mathrm{x}})+b,
\end{align}
where we defined the kernel $K(\boldsymbol{\mathrm{x}}_k,\boldsymbol{\mathrm{x}}_l)\coloneqq \phi(\boldsymbol{\mathrm{x}}_k)\cdot\phi(\boldsymbol{\mathrm{x}}_l)$, which is a compact function composed of the inner products between the mapped input data.

\begin{figure}[htpb]
\centering
\includegraphics[width=\linewidth]{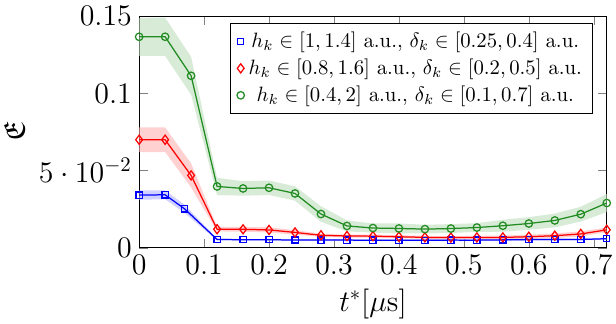}
\caption{Error in function of the measurement time $t^*$ for different intervals in which heights and widths of the potential barriers are varied. The reservoir parameters are fixed to $\Omega_{\rm glob}=7.5$ rad/$\mu$s, $V_{\rm NN}=1$ rad/$\mu$s, $\Delta_{\rm glob}=5$ rad/$\mu$s and $\Delta_{\rm loc}=-1.1$ rad/$\mu$s. The grid search is performed as in Fig.~\ref{fig:fig2}.}
\label{fig:err_intervals}
\end{figure}
Examples of kernel functions are the linear kernel $K_{\rm lin}(\boldsymbol{\mathrm{x}}_k,\boldsymbol{\mathrm{x}}_l)=\boldsymbol{\mathrm{x}}_k\cdot\boldsymbol{\mathrm{x}}_l$ and the Radial Basis Function (RBF) kernel $K_{\rm RBF}(\boldsymbol{\mathrm{x}}_k,\boldsymbol{\mathrm{x}}_l)=\textrm{exp}\left(-\gamma\|\boldsymbol{\mathrm{x}}_k-\boldsymbol{\mathrm{x}}_l\|^2\right)$. In the former case, no mapping is performed, while in the latter case mapping is performed in an infinite-dimensional feature space~\cite{pml1Book}. Other widely used kernel functions are polynomial kernels and sigmoid kernels. Together with the aforementioned classical techniques, quantum systems have been proposed to improve the performance of SVM~\cite{huang2021power}. In particular, the concept of Quantum Kernel (QK) has been introduced. The feature space is enlarged by mapping data points in the Hilbert space of a quantum system and the kernel matrix elements are given by the fidelities between the different quantum states that encode the data. Alternatively, a valid choice of kernel that intrinsically keeps trace of the properties of a quantum system are the Projected Quantum Kernels (PQK)~\cite{huang2021power,damore2024projected}; here, data are initially mapped into the Hilbert space of a quantum system and subsequently mapped back into a classical space. 

In our case, the input data are the potential features $\bm{\mathrm{v}}_k$, while the expected outputs are the density expansion coefficients $\{ u_k^{(\ell)}\}_{\ell=1}^{N_{\rm trunc}}$. We perform $\ell$ different SVRs, each with predicted output
\begin{equation}
    \tilde{u}_s^{(\ell)}=\tilde{u}^{(\ell)}(\boldsymbol{\mathrm{v}}_s)=\sum_{k \in \rm SV}\left(\alpha_k^{(\ell)} - \alpha_k^{(\ell)*}\right)K(\boldsymbol{\mathrm{v}}_k,\boldsymbol{\mathrm{v}}_s)+b^{(\ell)}.
\end{equation}
Regarding the kernel, we consider a particular class of PQK which is built as described in Sec.~\ref{sec:methods}. Input data $\boldsymbol{\mathrm{v}}_k$ are mapped in the measurement vector $\boldsymbol{\mathrm{m}}_k=\phi(\boldsymbol{\mathrm{v}}_k)$~\eqref{eq:measurement} which is a string of expectation values of observables extracted from a quantum reservoir. The PQK will be a linear kernel in $\boldsymbol{\mathrm{m}}_k$~\cite{kornjaca2024large}: 
\begin{align}
K_{\rm PQK}(\boldsymbol{\mathrm{v}}_k,\boldsymbol{\mathrm{v}}_l)&=\boldsymbol{\mathrm{m}}_k\cdot\boldsymbol{\mathrm{m}}_l \nonumber\\ & =  \sum_{j=0}^{L-1} \operatorname{Tr}\left[\rho_{t^*}(\bm{\mathrm{v}}_k)\sigma_j^z \right]\operatorname{Tr}\left[\rho_{t^*}(\bm{\mathrm{v}}_l)\sigma_j^z \right] \nonumber\\& + \sum_{i<j=0}^{L-1} \operatorname{Tr}\left[\rho_{t^*}(\bm{\mathrm{v}}_k)\sigma_i^z \sigma_j^z \right]\operatorname{Tr}\left[\rho_{t^*}(\bm{\mathrm{v}}_l)\sigma_i^z \sigma_i^z \right], 
\end{align}
where we wrote the expectation value $\braket{\ldots}_k=\braket{\psi_k(t^*)|\ldots|\psi_k(t^*)}$ with the general expression $\braket{\ldots}_k = \operatorname{Tr}\left[\rho_{t^*}(\bm{\mathrm{v}}_k) \ldots \right]$, $\rho_{t^*}(\bm{\mathrm{v}}_k)$ being the time evolved state of the reservoir with input data $\bm{\mathrm{v}}_k$ encoded in the Hamiltonian.

Results are compared with classical methods $K_{\rm lin}(\boldsymbol{\mathrm{v}}_k,\boldsymbol{\mathrm{v}}_l)=\boldsymbol{\mathrm{v}}_k\cdot\boldsymbol{\mathrm{v}}_l$, $K_{\rm RBF}(\boldsymbol{\mathrm{v}}_k,\boldsymbol{\mathrm{v}}_l)=\textrm{exp}\left(-\gamma\|\boldsymbol{\mathrm{v}}_k-\boldsymbol{\mathrm{v}}_l\|^2\right)$. In both cases, data are rescaled in such a way that the $\bm{\mathrm{v}}_k$ elements belong to the interval $[0,1]$.

\section{Additional properties of the error in the triple-well system}
\label{app:additional}
In this section, we analyze additional properties of the error that have not been covered in the main text. We refer to the triple-well potential problem and first study the scaling of the error with the number of training samples $N_{\rm train}$, consequently we analyze and its behavior for different values of the heights and widths of the barriers of the potential $V_k(x)$~\eqref{eq:potential}.

We first consider the scaling of the error, which is shown in Fig.~\ref{fig:err_scaling}; the number of hidden samples is kept fixed to $N_{\rm hidden}=20$. We work in a regime in which the PQK method performs well for $N_{\rm train}=20$. The error decreases for increasing $N_{\rm train}$ and also the gap between the error of the Linear kernel method with those of RBF and PQK methods appears to be more evident. For small values of $N_{\rm train}$, the average value of the PQK error is the smallest, while for increasing $N_{\rm train}$ it tends to converge to the one obtained with RBF kernel.    

Secondly, we consider the behavior of the error as a function of the measurement time in the case in which potential features are extracted in different intervals. In Fig.~\ref{fig:err_intervals}, we show the aforementioned behavior for three different choices of the intervals; $h_k$ compactly identifies the heights $h_{1,k}$ and $h_{2,k}$, while $\delta_k$ compactly identifies the widths $\delta_{1,k}$ and $\delta_{2,k}$. The red curve corresponds to the same range of potential features considered in the main text and defined in Appendix~\ref{app:one_dim}.  It is clear that for larger intervals the densities associated to different samples are less correlated and so the learning process is harder. This qualitative expectation is confirmed by our results that show smaller error for smaller intervals of the potential parameters. We also observe that the time at which the error drops is the same for all the three cases.

\end{document}